# Enhancing interfacial thermal conductance of Si/PVDF by strengthening atomic couplings

Zhicheng Zong[1], Shichen Deng[1], Yangjun Qin[1], Xiao Wan[1], Jiahong Zhan[1], Dengke Ma[2], Nuo Yang[1*]

1. School of Energy and Power Engineering, Huazhong University of Science and Technology, Wuhan 430074, China.
2. Phonon Engineering Research Center of Jiangsu Province, Center for Quantum Transport and Thermal Energy Science, Institute of Physics and Interdisciplinary Science, School of Physics and Technology, Nanjing Normal University, Nanjing 210023, China

*Corresponding email: nuo@hust.edu.cn (N.Y)

# Abstract

The thermal transport across inorganic/organic interfaces attracts interest for both academic and industry due to its widely applications in flexible electronics etc. Here, the interfacial thermal conductance of inorganic/organic interfaces consisting of silicon and polyvinylidene fluoride is systematically investigated by molecular dynamics simulations. Interestingly, it is demonstrated that a modified silicon surface with hydroxyl groups can drastically enhance the conductance by 698%. These results are elucidated based on interfacial couplings and lattice dynamics insights. This study not only provides feasible strategies to effectively modulate the interfacial thermal conductance of inorganic/organic interfaces but also deepens the understanding of the fundamental physics underlying phonon transport across interfaces.

# Introduction

Inorganic/organic interfaces widely exist in flexible electronics, optoelectronics, photovoltaic, thermoelectric, etc[1-6]. Devices built from hybrid inorganic and organic materials can be designed to combine the complementary strengths of distinct materials[7, 8]. PVDF is often referred to as a favorite polymer from the family of organic materials due to its excellent piezoelectric properties, thermal stability, and mechanical strength[9, 10]. In the application of PVDF to wearable flexible devices and batteries, the silicon/polyvinylidene fluoride (Si/PVDF) interface is commonly utilized and representative[11-13].

The study of interfacial thermal conductance (ITC) is urgent because the high density of interfaces blocks heat dissipation and takes thermal damage to devices[14-17]. Especially in nanodevices and structures containing high-density interfacial structures, the characteristic size has been reduced to the magnitude of the average free path of energy carriers, and the thermal resistance of atomic-scale interfacial structures cannot be ignored. Hence, it is imperative to gain a comprehensive understanding of the atomic-scale interfacial thermal transport mechanism and enhance the interfacial thermal conductance accordingly[18-21].

The modification of surface structures is an effective approach to enhance ITC[21]. Currently, there are various approaches available to enhance ITC, including the utilization of interface defects[22-24], interlayers[25-28] and surface chemical treatment[29-35]. Among these approaches, surface chemical treatment has been demonstrated through different experiments as a viable method for modifying surface properties[35-38]. This chemical treatment exhibits significant potential in improving the atomic-scale forces and consequently enhancing the ITC at the interface. Therefore, in the case of Si/PVDF interfaces, modifying the silicon surface with hydroxyl groups[39, 40] has the potential to enhance ITC.

In this work, the interfacial thermal conductance (ITC) of Si/PVDF is enhanced by modifying the surface with hydroxyl. The study is performed by non-equilibrium molecular dynamics (NEMD) simulations. Firstly, Si/PVDF interface with intrinsic and modified Si with hydroxyl group are constructed. Secondly, the intrinsic and modified dependence of ITC is studied. Thirdly, an analysis of atomic couplings and lattice dynamics is carried out to explain the difference between intrinsic and modified Si surfaces on ITC. Lastly, the effect of temperature difference of Si/PVDF interface on ITC is studied and the outlook of future research is proposed.

## Structure and method

The simulation cells of different Si/PVDF were formed and shown in Fig. 1(a)-(f). In order to investigate the roles of hydroxyl groups (OH) in interfacial heat transfer, the intrinsic silicon (In-Si) and modified silicon (Mod-Si) surfaces facing PVDF are structured. Furthermore, poled-PVDF (P-PVDF), unpoled-PVDF (U-PVDF), and amorphous PVDF (A-PVDF) are structured to explore the effects of modulation on ITC. All the details of the structures are provided in Supplementary Material (SM) 1.

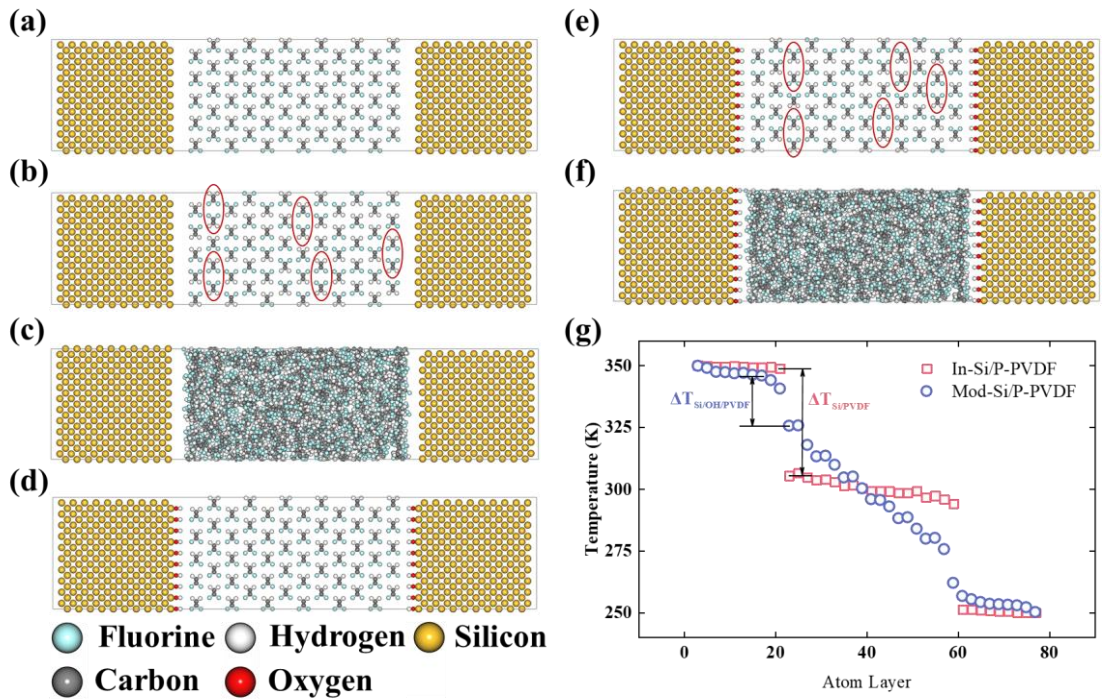

Fig .1. Schematics of the structure of (a) intrinsic silicon/poled PVDF (In-Si/P-PVDF) interface, (b) intrinsic silicon/unpoled PVDF (In-Si/U-PVDF) interface, (c) intrinsic silicon/amorphous PVDF (In-Si/A-PVDF) interface, (d) modified silicon/poled PVDF (Mod-Si/P-PVDF) interface, (e) modified silicon/unpoled PVDF (Mod-Si/U-PVDF) interface and (f) modified silicon/amorphous PVDF (Mod-Si/A-PVDF) interface. (g) The temperature difference of the structure along with the z direction calculated by NEMD is recorded at 100K temperature difference. The thermal resistance of the intermediate layer OH is taken into account in the interface thermal resistance.

MD simulations were implemented using the Large-scale Atomic/Molecular Massively Parallel Simulator (LAMMPS)[41] with the force field of PCFF/Lennard-Jones/Tersoff[42,43]. Prior to the calculation of ITC, the structures underwent a full relaxation process. Then the temperature difference of the structure are recorded at 100K temperature difference, as shown in Fig. 1(g). Once the temperature difference and heat flux (SM 4) at the interface are determined, the thermal conductance of the interface can be calculated accordingly. Notably, when considering Mod-Si/PVDF interfaces, the thermal resistance of the intermediate layer OH is taken into account in the interface thermal resistance ($R_{\text{inter}} = R_{\text{Si-OH}} + R_{\text{Si-PVDF}}$). More details of the simulation are provided in the SM, from SM2 to SM6.

# Results and discussion

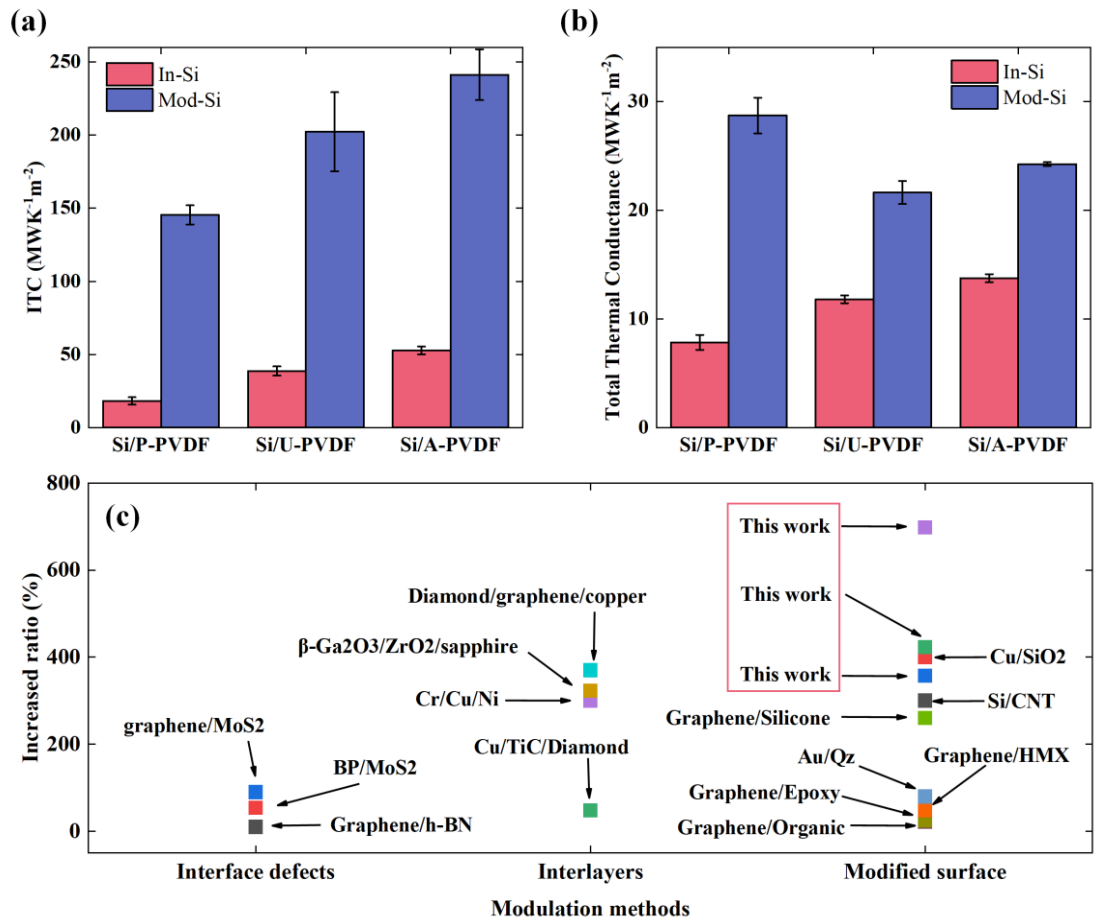

Fig .2. (a) ITC of PVDF/Si interface: For Si/P-PVDF, Si/U-PVDF, and Si/A-PVDF interface, through modification of Si surface ITC can be increased by 698%, 423%, and 357%. (b) Total conductance of composite structures: For Si/P-PVDF, Si/U-PVDF, and Si/A-PVDF composite structures, through modification of Si surface total conductance of composite structures can be increased by 266%, 83%, and 76%. (c) Increased ratio versus other modulation methods (Graphene/h-BN[22], BP-MoS2[23], Graphene/MoS2[24], Cu/TiC/Diamond[25], Cr/Cu/Ni[26], β-Ga2O3/ZrO2/sapphire[27], Diamond/Graphene/copper[28], graphene/organic[29], graphene/epoxy[30], graphene/HMX[31], Au/Qz[32], graphene/silicone[33], Si/CNT[34] and Cu/SiO2[35])

The ITCs of different Si/PVDF interfaces are shown in Fig. 2. The ITC of modified Si surface is much larger, up to 698% than that of intrinsic case. By employing NEMD simulation, the ITCs of the six different simulation cells above is determined at 300K. Amongst the same silicon surfaces, it is observed that A-PVDF exhibits a higher ITC in comparison to U-PVDF, whereas U-PVDF demonstrates a higher ITC than P-PVDF.

The observed difference in ITC can primarily be attributed to the higher level of disorder in A-PVDF, which results in a stronger coupling to the interface. On the other hand, P-PVDF is characterized as the weakest among the PVDF, contributing to its lowest ITC. Besides, it is observed that the modifications applied to the Si surface using the properties of the hydroxyl group led to a substantial enhancement in the ITC. Compared with In-Si/PVDF interface, the ITC values of Mod-Si/P-PVDF, Mod-Si/U-PVDF, and Mod-Si/A-PVDF interfaces are increased by 698%, 423%, and 357%, respectively. More importantly, for Mod-Si/P-PVDF interfaces, the ITC experienced a substantial increase from 18.23 to 145.44 $MWK^{-1}m^{-1}$ (693%) when compared to the In-Si/P-PVDF interface. As plotted in Fig. 3 (c), the enhancement of this work is compared with other methods, such as interface defects[22-24] and interlayer[25-28].

Besides, the thermal conductance of the whole composite structure is studied, as shown in Fig. 2(b). Through modification of the Si surface, the total thermal conductance of composite structures of Si/P-PVDF, Si/U-PVDF, and Si/A-PVDF can be increased by 266%, 83%, and 76%，res. Notably, the thermal conductivity of Si and PVDF remained unchanged before and after the modification of the Si surface (SM5). The primary determinant of thermal conductance variation in the composite system is attributed to alterations in ITC. On one hand, the result indicates that by modifying the Si surface with hydroxy groups, the thermal transport of the Si/PVDF interface and total composite structures can both be largely increased. On the other hand, the results also show that thermal transport across the interface plays a great role at the nanoscale.

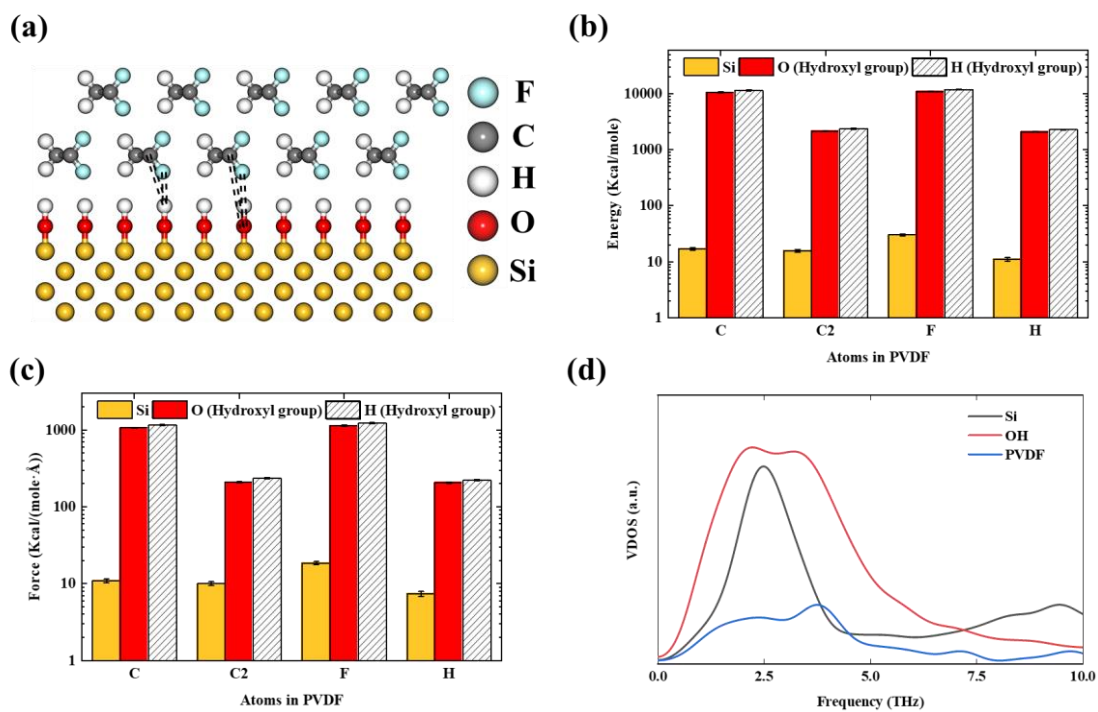

Fig .3. (a) Schematic of atomic force in Mod-Si/P-PVDF composite structures. Strong atomic force can make organic molecules closer to the interface, further increasing the interface binding energy. (b) Total energy and (c) force interaction between different groups of atoms. The hydroxyl group on the modified silicon surface has a stronger interaction with the atoms in PVDF, which largely enhanced ITC by strong interfacial coupling. (d) The vibrational density of states (VDOS) of Mod-Si/P-PVDF composite structures indicates that the coincidence area of PVDF and Si (16.1%) is slightly smaller than that of PVDF and OH (16.4%).

To elucidate the mechanism behind ITC changes, the interfacial atomic structures are plotted in Fig. 3(a). The local structure of the interface shows the modified silicon surface and the atoms in the PVDF at the interface. After the modification of the Si surface, the interaction at the interface undergoes a transformation from a van der Waals force to a coulomb force. The findings reveal that the atoms in PVDF exert a strong coulomb force on the hydroxyl group located on the surface of the modified silicon, as shown by the black dotted line in the Fig. 3(a). Additionally, all other relaxed structures are displayed in Fig. S1.

Generally, a stronger atomic coupling can take a more efficient heat conduction. This is primarily because stronger interfacial couplings provide multiple paths for thermal transport, which can exert a significant influence on the transmission probabilities across interfaces[44]. Then total energy and force interaction between different groups of atoms is quantitatively compared in Fig. 3(b) and (c). The observations indicate that the hydroxyl group on the modified silicon surface exhibits a stronger interaction with the atoms in PVDF, which make organic molecules closer to the interface, thereby further increasing the interface binding energy and the transmission probabilities across the interface[37, 44]. As a result, strong atomic coupling contributes significantly to the enhanced ITC.

To further understand the mechanism, the vibrational density of states (VDOS) is calculated and analyzed qualitatively. For VDOS of Mod-Si/P-PVDF composite structures, the coincidence area of PVDF and Si (16.1%) is marginally smaller than that of PVDF and OH (16.4%), as shown in Fig. 3(c). Based on the theoretical model[16, 45], the hydroxyl group situated on the surface of modified silicon displays a higher degree of matching with PVDF, resulting in an increased supply of thermal transport channels and thus enhanced ITC.

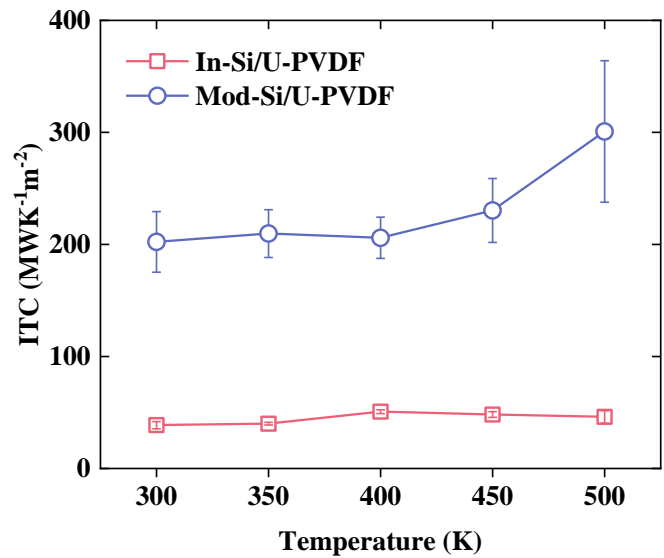

Fig .4. ITC of In-Si/U-PVDF and Mod-Si/U-PVDF interface varies with different temperatures. High-frequency modes stimulated at high temperature can be transmitted

across the interface through evanescent modes, facilitated by the inelastic three-phonon scattering process, resulting in the enhancement of ITC.

Further, the temperature dependence of ITC is studied from 300K to 500K (Fig.4). The results demonstrate a positive correlation between ITC and temperature, indicating that an increase in temperature leads to a corresponding increase in ITC, particularly for the Mod-Si/U-PVDF interface. The increase in temperature promotes the occurrence of multi-phonon processes at the interface, thus creating new pathways for heat transport[24,46]. The In-Si/U-PVDF interface exhibits a weak interaction that hinders the transportation of high-frequency phonons. Conversely, the Mod-Si/U-PVDF interface demonstrates a strong interaction, which facilitates the transportation of high-frequency phonons. Besides, the inelastic high order phonon scattering process also results in the enhancement of ITC[16]. The introduction of hydroxyl groups on the Si surface introduces numerous phonon modes at higher frequencies (as shown in SM6), which contribute to the increased ITC variations at interface as the increase of temperature.

It is noted that the system does not melt at 500 K which surpasses the melting point of amorphous PVDF(450 K)[47]. The melting points of polymers demonstrate an increase in chain length and lamellar crystal thickness[48-50]. In simulations of this Fig 4, the studied structures of PVDF are crystalline-like, and periodic boundary conditions are applied, which simulates an infinite chain length and lamellar crystal thickness. Therefore, the melting point in systems is higher than 450 K, and the system does not melt at 500 K. These findings are consistent with similar phenomena reported in previous studies conducted by MD simulations, such as PE[51], PEO[52], PEDOT[53], PVDF[47] and nylon[54].

# Conclusion

In summary, it is investigated that the modification effect on the thermal transport across Si/PVDF interface by molecular dynamics simulations. The Si/PVDF interfaces are modified with hydroxyl groups, where atomic couplings between Si and PVDS are strengthening. The interfacial thermal conductance across modified interfaces is significant enhanced compared with that of the intrinsic interface. The value of ITC across the Mod-Si/P-PVDF enhanced from 18.23 to 145.44 MWK$^{-1}$m$^{-1}$ (693%). The interfacial couplings analysis reveals that the increased interfacial bonding leads to a dramatic increase in ITC. Then, the VDOS results show that the modification of the Si surface leads to more phonon transport across the interface. Moreover, the ITC of Mod-Si/U-PVDF and In-Si/U-PVDF simulated in different temperatures are calculated, and the results show that ITC could be enhanced with the increment of temperature. It is believed that fine-tuning the process of modified surfaces can achieve an optimal structure of inorganic/organic interfaces with both higher interfacial adhesion and better mechanical properties. This study illustrates an effective avenue to tune thermal transport performance across inorganic/organic interfaces for thermal management and energy efficiency applications.

## Conflict of interest

The authors have no conflicts to disclose.

## Acknowledgement

This work is sponsored by the National Key Research and Development Project of China No. 2018YFE0127800. The authors thank the National Supercomputing Center in Tianjin (NSCC-TJ) and the China Scientific Computing Grid (ScGrid) for providing assistance in computations.